\documentclass[%
 reprint,
 amsmath,amssymb,
 aps,
]{revtex4-1}

\usepackage{graphicx}
\usepackage{dcolumn}
\usepackage{bm}
\usepackage[english]{babel}
\usepackage{fancyhdr}
\usepackage{amssymb}
\usepackage{amsmath}
\usepackage{color}
\usepackage[bottom]{footmisc}

\begin{document}

\preprint{APS/123-QED}

\title{Reexamining a Geometric Theory of Biological Growth}

\author{Dominic Gastaldo}
\author{David Silverstein}


\date{\today}

\begin{abstract}
A first principles approach to the theoretical description of the development of biological forms, from a fertilized egg to a functioning embryo, remains a central challenge to applied physics and theoretical biology. Rather than refer to principles of self-organization and non-equilibrium statistical mechanics to describe a developing embryo from its active cellular constituents, a purely geometric theory is constructed that references the properties of the ambient space that the embryo occupies. In 1975 the Fields laureate Ren\'{e} Thom developed a system of techniques and local dynamical models that are capable of reconstructing the local dynamic of an embryo at each new growth event of the system. Each new growth event (the development of a limb, for example) is a topological change in the dynamic of the system that can be classified only according to the properties of space. The local models can be non-conservative flows with robust attractor behavior that serve as organizing centers for systems development. Hamiltonian flows can also be considered with novel, self-reproducing vague attractor behavior. The set of growth events become an unfolding space related to the differentiable manifold of states. The set of growth events, which Thom refers to as the catastrophe set, has special algebraic properties which permit these models to be low dimensional--the local model contains few parameters. We present Thom's work as a research program outlining a framework for the construction of these local models, and, ultimately, the synthesis of these models into a full theoretical description of a developing biological organism. We give examples of the application of selected models to key growth events in the process of gastrulation.

\end{abstract}

\pacs{Valid PACS appear here}
\maketitle


\section{\label{sec:level1}Introduction}

 In 1973, Ren\'{e} Thom published the first French edition of {\it Structural Stability and Morphogenesis: An Outline of a General Theory of Models}, a monograph outlining a research program to reconstruct the dynamical equations that govern the geometric change in form of a developing embryo. Thom presents a deeply philosophical meditation on the geometric principles of stability in our universe. Rather than investigate biological structure as a system composed of parts that have function, Thom articulates the view of a geometer as he constructs the framework for a topologically centric theory.  This theory formulates complex biological change as derivative properties of space. Spacetime naturally hosts systems that themselves have a dynamic that describes their evolution, and the topological changes in the dynamic qualitatively change its behavior, therefore the behavior of the system. These dynamics may be transient; changes in the environment, perturbations and thermodynamic fluctuations may disrupt a system and cause it to change how it behaves. The stable ways that a system can change its dynamic are deeply related to the geometry of space. Thom studies local differential models that depend on the most general, stable way a dynamic can topologically change its behavior.  These are the so-called universal unfoldings of the dynamic.  Thom develops a research plan around selected differential equations that depend on the catastrophe bifurcations in spacetime. Each developmental change in form is associated with a catastrophe bifurcation. Thom's program then becomes the study of structurally stable bifurcations that locally determine a system of differential equations describing each morphogenic event.

\section{\label{sec:level1}Theoretical formulation}

Thom's method can be summarized in a general mathematical framework. The general mathematical structure will take the form of a three-tuple $(M,X,K)$, where $M$ is the state space differentiable manifold, $X$ is the vector field dynamic encoding the solution to a system of differential equations governing the time evolution of the embryo, and $K$ is the algebraic set of catastrophe bifurcations. The state space $M$ is a finite dimensional space containing position and momenta of the cells that make up an embryo. Beyond the scope of Thom's framework, one can consider a sequence of manifolds $(M^{2}, M^{4},...,M^{2^{n}})$, for example, that each preserve the local dynamic of a particular growth process. Each member of $K$ represents a morphogenesis of the system. Locally, $K$ depends on the dynamic $X$. This permits $X$ to be locally reconstructed by the elements of $K$.

The catastrophe set $K$ has special algebraic properties that produce models with few parameters. The algebraic properties of $K$ also permit $X$ to be locally reconstructed even when all the parameters of $X$ are not directly known. The possible catastrophe bifurcations of a dynamical system can be classified according to the dimension and topology of the space that the embryo occupies.  The elements of $K$ are universal unfoldings of the singularities in $X$. The singularities in $X$ encode each morphogenetic event. Thom explores the ultimate consequence of this by considering a global model of an organism as the universal unfolding of its egg. We examine each local model as determined by the universal unfoldings of $X$ that make up $K$.

Each local differential model is determined by the elements of $K$. The analytic structure of $K$ determines the structural stability of the dynamic. If $K$ is everywhere dense, then $X$ is structurally stable. It's also clear that if $K$ is non-empty, then the dynamic must be non-conservative. However, we present how local Hamiltonian models can still be used. Thom experiments with conservative, Hamiltonian systems with vague attractors. Some models have very interesting self-replicating features. The gradient dynamic implies a Riemannian metric $g_{\mu \nu}$ equipped on the manifold $M$. This leads to the possibility of making quantitative an otherwise qualitative theory.  Finally, certain phenomenological principles need to be mentioned. A thermodynamical coupling may need to be defined between dynamical systems. This will define a measure-theoretic entropy and permit the two systems to come into thermal equilibrium. Also, a convention inspired from thermodynamics needs to be adopted. If two attractors are in competition, the attractor at the lowest potential is the attractor that determines the local dynamic. The bifurcation of a degenerate critical point can result in a shock wave as the critical points change their relative positions. The shock waves resulting from the emergence or annihilation of critical points becomes embryological growth waves. These phenomenological principles can be used together with the applied mathematical model to develop robust descriptions of local growth.

\subsection{\label{sec:level1}Topological Classifications}

This section, topological classifications, was first published in Thom's original {\it Structural Stability and Morphogenesis: An Outline of a General Theory of Models} (chapter 3). We have here simplified and reorganized Thom's original classifications to elucidate meaning to the reader. In this section, we follow Thom's outline for constructing a family of structurally stable unfoldings of degenerate critical points.  We will begin by looking at the specific topological classifications that are needed to satisfy the conditions for the universal unfolding.  The universal unfolding will define the morphogenic effects that will allow one to set up a general theory of models.  Therefore it is crucial for one to realize when structural stability enters the theory.  We will begin by examining a theory which is going to have the following homeomorphisms.  Let us take a continuous family of geometric objects $E_s$ where $s$ is a point within a space of parameters $S$.  Here $S$ can be represented as a Euclidean space or a differential manifold that is either finite or infinite-dimensional.  We can then do the same with a family of geometric objects $E_t$ where $t$ is a point in $T$, with $T$ once again representing what can be a Euclidean space or a differential manifold that is either finite or infinite-dimensional.  If the points $s$ and $t$ are sufficiently close to one another, then the corresponds points or objects will have the same topological form and can be classified as structurally stable.  If such points are structurally stable, they form an open subset of structurally stable points classified as generic points.  The corresponding open subset has a compliment $K$ which is defined as the set of bifurcation points.  From here we can take a system of polynomial equations \begin{equation}
 P_j(X_i, S_k)=0
\end{equation}
Here $X_i$ and $S_k$ are coordinates in Euclidean space $\mathbb{R}^{n}$ and $S$.  Taking $S=(S_k)$ being fixed we get an algebraic set $E_s$ in $\mathbb{R}^{n}$ which contains a set of solutions of the equation, where the coefficients could be both real or imaginary.  With the given points $s$ and $t$ being in the same component of $S-K$, our sets $E_s$ and $E_t$ are homeomorphic and can be invariant under continuous deformations.  Such homeomorphic deformations play a critical role in defining the properties of the universal unfolding.  Let us now take two differential manifolds labeled $X$ and $Y$, with $X$ being compact.  This allows us to write down the function space $L(X,Y)$ of our differential maps from $X$ to $Y$.  Such a differential mapping is denoted as the $C^m$ topology where the differential maps are defined on a chart by the following metric
\begin{equation}
d(f,g)=\sum_{|k|=0}^m\bigg| \frac{\partial^{|k|}}{\partial x^k}(f-g)\bigg|
\end{equation}
Here we see that the infinite dimensional manifold is a Banach manifold due to the labeling being done on a Banach space.  Yet we specifically see that such a labeling on the Banach space is a local labeling which sets the stage to allow us to have a geometric theory of local models.  With that said, the function space $L(X,Y$) containing our two manifolds plays the role of the parameter space $S$.  If we were to take each point $F$ in the parameter space $S=L(X,Y)$, we would see that this is a differentiable map $X \rightarrow Y$.  We know if we were to take two maps $f$ and $g$ pertaining to $X \rightarrow Y$ that they have the same topology type under homeomorphism.  With both of these being homeomorphic we can write down the two homeomorphisms $h: X \rightarrow X$ and $k: Y \rightarrow Y$.  Because of this setup we now have in $L(X,Y)$ a subspace $K$ being our closed bifurcation set.  We now can have any two maps $f$ and $g$ of $L(X,Y)$ which in turns give us a connected component $L-K$.  If our maps $f$ and $g$ that belong to $L(X,Y)$ in turn belong to the same connected component $L-K$, they will have the same topology type.   It is important to note that the bifurcation set $K$ has a stratified structure containing regular points that form an everywhere-dense open set in $K$.  When locally studying the stratum of the bifurcation set of codimension $K$, we will see that such an analysis is going to play a key role in defining the universal unfolding.  To give a good idea of a stratified map, let us take the following canonical map
\begin{equation}
F\colon X\times L(X,Y)\rightarrow Y \times L(X,Y)\rightarrow L(X,Y)
\end{equation}
which is defined by
\begin{equation}
(x,f)\rightarrow(y=f(x),f)\rightarrow f
\end{equation}
This map gives the existence of a closed subspace $H$ of infinite codimension in $L(X,Y)$.  Now the function space $L(X,Y)$ can now be written as $L(X,Y)-H$, which gives the existence of a stratified map.  However, we can recall that our $L(X,Y)$ has the subspace of the closed bifurcation set $K$.  Therefore the stratification $L(X,Y)-H$ is defined by the bifurcation subset $K$.  Since the stratification of $K$ contains $H$ pertaining to the function space $L(X,Y)$, we can rewrite our stratification of $L(X,Y)-H$ as $L(X,Y)-K$.  Such maps are structurally stable maps and can therefore be labeled as generic maps.  Due to the maps in $L(X,Y)-K$ being generic, $L(X,Y)-H$ can then be embedded in what Thom calls a $q-parameter$ family where the map $X\times \mathbb{R}^{q} \rightarrow Y \times \mathbb{R}^{q}$ will be defined as generic. Perturbations leave the parameter space $\mathbb{R}^{q}$ invariant.  To see the proof of this, we first need to study the problem locally.  We can begin by taking the following germ $g\colon \mathbb{R}^{m} \rightarrow \mathbb{R}^{n}$ of a differential map which has the following origin $O$. Such a germ allows one to send the origin to $O$ in $\mathbb{R}^{m}$ to the origin $O_1$ in $\mathbb{R}^{n}$ which gives an expansion of the map.  With such an expansion it will be important to go over the jet space that defines the expansion.  We can take the following jet space $z$ of order $r$ and an integer $r+p$ with $p$ depending on parameters $m$, $n$ and $r$.  We can then extend the jet to order $r+p$ which allows one to determine the local topology of the germ.  Such a property is contained within the algebraic subset which is the bifurcation set $K$.  By taking the expansion of the origin for $O$ to $O_1$ pertaining to the germ with such proprieties in the bifurcation set allows one to take the stratification of the jet space $z$.  Just like before, such a stratification will give rise to a structurally stable jet which will once again be invariant under perturbations.  While we have been carefully examining the topological classifications, we should now turn on our attention to what actually controls morphogenesis.  With that said, we should now focus on a more detailed example of a bifurcation that relates to both a stratified jet space and the expansion of the origin.  With the expansion and stratification of a jet space, we can easily consider a differentiable function that is zero at the origin, which will have the following Taylor series expansion

First order:
\begin{equation}
f_{1} = a_1x + b_1y
\end{equation}

Second order:
\begin{equation}
f_{2} = a_1x + b_1y + a_2x^2 + 2b_2xy + c_2y^2
\end{equation}

Third Order:
\begin{equation}
f = a_1x + b_1y + a_2x^2 + 2b_2xy + c_2y^2 + a_3x^3 + 3b_3x^2y + 3c_3xy^2 + d_3y^3
\end{equation}
Now lets write down the successive space of coefficients $J^1(a_1,b_1), \: J^2(a_1, b_1, a_2, b_2, c_2), \: J^3(a_1,...,c_3, d_3)$.  To first order in $J^1$ we have the following topological classification where either both coefficients $a_1$, $b_1$ are not equal to zero or $a_1=b_1=0$.  By taking a careful look at the first case with $a_1$ and $b_1$ being nonzero at the origin, and due to the equations being expressed locally, we can easily use the implicit function theorem.  By using the implicit function theorem as noted by Thom, we see that we can locally change to curvlinear coordinates which will give us a linear function, and will in turn remain invariant regardless of higher coefficients.  If we look at the second case where $a_1=b_1=0$; very little can be said regarding the topological nature of the function due to everything being centered at the singularity.  Yet we have indeed seen such topological properties go through such processes regarding the expansions of origins due the expansion and stratification of the jet spaces.  We have also seen that the expansion of such jet spaces satisfies the properties of a bifurcation set.  Hence, we can now restrict our attention to the expansion of a singularity of finite codimension.  Begin by taking the local maps $f$ and $g$ being two local maps in Euclidean space $\mathbb{R}^n\rightarrow \mathbb{R}^p$, which map the origin $O$ on $\mathbb{R}^n$ the origin $O_1$ in $\mathbb{R}^p$.  These maps can have the same local jet of order $r$ at $O$, and we can in turn have the set of such jets $J^r(n,p)$ forming a vector space containing the coordinates of the coefficients of the Taylor series expansion.  Due to this having a bifurcation subset, we can write $J^r(n,p)$ as $J^r(n,p)-K^r$, and the maps $f$ and $g$ will locally have the same topology type.  Therefore there will be a homeomorphism $h$ and $h_1$.  We can then write out the singularities as $h(O)=O_1,\:h_1(O_1)=O_1$.  As stated before, because the bifurcation set $K$ is a subspace, there is going to a corresponding finite codimension.  Like before, we can write our codimension as $q$ due to it being able to be embedded in a map $F\colon\mathbb{R}^n\times \mathbb{R}^q\rightarrow \mathbb{R}^p\times \mathbb{R}^q$, is again defined as structurally stable, being invariant under perturbations.  Due to these all being apart of the $C^m$ topology which extends to the parameter q, we have the expansions of our singularities extending from $m$ to $q$.  All these parameters have been shown to invariant under perturbations, hence, it seems natural that we can write out a universal family pertaining to the expansion of the singularities in the form of  $\mathbb{R}^m \rightarrow \mathbb{R}^q$.  This allows us to examine the unfolding of the singularities that pertain to the universal map  $\mathbb{R}^m \rightarrow \mathbb{R}^q$, which satisfies the properties of the bifurcation set $K$.  Such an unfolding is defined as the universal unfolding, and from here we can clearly illustrate an example of the universal unfolding.  From the Taylor series, we have shown that the perturbations occur at third-order.  Therefore, we will study the universal unfolding of $y=x^3$.  Every deformation of $y=x^3$ is a differentiable function $y=f(x,v)$ depending on $m$ parameters denoted globally by $v$, where $f(x,0)=x^3$.  We write
\begin{equation}
F(x,y,v)=f(x,v)-y
\end{equation}
Then, at the origin $x=y=v=0$\: of\: $(x,y,v)-space$, $F$ satisfies
\begin{equation}
F(0,0,0)=\frac{\partial}{\partial x}F(0,0,0)=\frac{\partial^2}{\partial x^2}F(0,0,0)=0, \frac{\partial^3}{\partial x^3}F(0,0,0)\neq 0
\end{equation}
Due to the analytic nature of this function, it was pointed out by Thom, that we can easily apply Weierstrass preparation theorem and get the following
\begin{equation}
F(x,y,v)=h(x,y,v)[x^3-3a(y,v)x^2+b(y,v)x+c(y,v)]
\end{equation}
Here the functions $h,a,b $ and $c$ are differentiable functions that satisfy $h(0,0,0)\neq 0 \:\:\:\: a(0,0)=b(0,0)=c(0,0)=0$  and \begin{equation}\frac{\partial c}{\partial y}(0,0)=\frac{\partial}{\partial y}\bigg[\frac{F(x,y,v)}{h(x,y,z)}\bigg]_{{x={y=z}}} =\frac{1}{h(0,0,0)}\neq 0
\end{equation}
By taking the Taylor expansion, this allowed us to show how the universal unfolding occurs at third order, which helps further illuminate in addition to show the conditions needed to satisfy the universal unfolding.  Now that the topological conditions of the universal unfolding have been satisfied, we are now able to move onto the next step which will be defining and classifying the catastrophes.

\subsection{\label{sec:level1} Local Dynamical Models of Morphogenesis}

In this section, we will examine and build upon Thom's reconstruction of a local gradient dynamic. Thom classifies elementary catastrophes in $\mathbb{R}^4$. Elementary catastrophes are bifurcations of a gradient dynamic $\dot x = -\nabla V(x)$. To fully realize Thom's program, catastrophes of more complex flows need to be studied along with their associated morphologies. This may be very challenging, and we study only the gradient dynamic in detail. These elementary catastrophes produce a bifurcation that results in a structurally stable dynamic associated with a morphology. Each point where both the derivative and the second derivative vanish can be universally unfolded into a catastrophe that represents a structurally stable dynamic (i.e. stable to perturbations). Each elementary catastrophe has a codimention, which is the number of additional parameters that are added on to the original function. The bifurcation then exists in the space defined by the additional parameters. Elementary catastrophes in $\mathbb{R}^4$ are associated with a morphology. Elementary catastrophes map onto a local gradient dynamic.  

We wish to construct models that are locally equivalent to the global dynamic $X$ within the neighborhood of each point $x \in$ K. These models will depend on the algebraic properties of the catastrophe set, and both the spacial parameters and the parameters in the universal unfolding will depend on properties of the critical point and its universal unfolding. Though each model can be defined on $\mathbb{R}^n$, the relevant models reduce down to the corank of the critical point, either one or two dimensions. These are the minimum dimensions necessary to describe the {\it topology of the underlying dynamic}. There is nothing preventing the dynamic from being embedded in $\mathbb{R}^3$, and indeed, this is appropriate to endow these models with further quantitative predictive power.

\subsubsection{The Gradient Model}

We first consider the gradient model, and we examine the classifications of elementary catastrophes associated with a local gradient dynamic.

\subsubsection{Fold}

The fold is the simplest non trivial catastrophe. It has codimension one, and it is the universal unfolding of $V(x) = \frac{1}{3}x^3$.

\begin{equation}
V(u, x)=  \frac{1}{3}x^3 + ux
\end{equation}

The fold has no morphogenetic effect, and it can't generate shock waves.

\subsubsection{Cusp}

The cusp catastrophe is defined as

\begin{equation}
V(u, v, x) = \frac{1}{4}x^4+ \frac{1}{2}ux^2 + vx
\end{equation}

The gradient model determines the time evolution equation

\[
\frac{dx}{dt} = -\frac{d}{dx} (\frac{1}{4}x^4+ \frac{1}{2}ux^2 + vx)
\]

\[
\frac{dx}{dt} = -x^3 - ux - v
\]

This differential equation undergoes a cusp catastrophe for the variation of the parameters $(u,v)$. It has a set of fixed points $-x^3 - ux - v = 0$.

This elementary catastrophe becomes the structurally stable cuspidal of caustics.

\subsubsection{Swallow tail}

The swallow tail catastrophe can produce growth waves as the stable and unstable regions of its bifurcation set coincide. In optics, it can be interpreted as the formation of caustics. In embryological development it becomes the blastopore fold. During gastrulation, the embryo folds in through invagination, and the swallow tail is the catastrophe that captures the local dynamic of the formation of a stable fold.

\begin{figure}[h!]
  \includegraphics[width=1.0\columnwidth]{invaginate.pdf}
  \caption{The invagination of Drosophila. This begins the gastrulation process and the formation of the digestive system. \footnote{[image credit: Conte V, Ulrich F, Baum B, Muñoz J, Veldhuis J, Brodland W, Miodownik M]}}
\end{figure}

The potential is the universal unfolding of $V(x) = \frac{1}{5}x^5$

\begin{equation}
V(u,v,w,x) = \frac{1}{5}x^5 + \frac{1}{3}ux^3 + \frac{1}{2}vx^2 + wx
\end{equation}

With local model

\[
\dot x = -x^4 - ux^2 - vx - w
\]

This implies, if indeed the swallow tail models invagination as Thom suggests, that invagination is a one dimensional topological dynamical system.

\subsubsection{Hyperbolic umbilic}

The hyperbolic umbilic is given by the unfolding of $V(x,y) = x^3 + y^3$.

\begin{equation}
V(u,v,w,x,y)= x^{3} - y^{3} + wxy - ux - vy
\end{equation}

Thom doesn't go into detail about the embryological effect of the Hyperbolic umbilic. Rather, he relates it to the geometry of the breaking of waves.

\subsubsection{Elliptic umbilic}

The elliptic umbilic can be used to locally model hair, spikes or protrusions
\begin{equation}
V(u,v,w,x,y)= x^{3} - 3xy^{2} + w(x^{2}+y^{2}) - ux - vy
\end{equation}

This model will produce the system of first order non-linear equations

\[
\dot x =  -3x^{2} + 3y^{2} - 2wx + u
\]

\[
\dot y =  6xy - 2yw + v
\]

\subsubsection{Butterfly}

We can unfold a $V(x) = x^6$ potential to describe the elementary catastrophe resulting in an exfoliating blister. This is known as the butterfly catastrophe.
\begin{equation}
V(u,v,w,s,x) = \frac{1}{6}x^6 + \frac{1}{4}sx^4 + \frac{1}{3}ux^3 + \frac{1}{2}vx^2 + wy
\end{equation}

The butterfly catastrophe can generate an exfoliating blister.

\subsubsection{Parabolic umbilic}

The parabolic umbilic describes a mushroom morphology. Its potential is $V(x,y) = x^2 y + \frac{y^4}{4}$, and its universal unfolding is

\begin{equation}
V(u,v,w,s,x,y) = x^2 y + \frac{y^4}{4} + sx^{2} + wy^{2} -ux -vy
\end{equation}

The equations of motion describing the local dynamic are then

\[
\dot x =  2x(y + s) - u
\]

\[
\dot y =  x^2 + y^3 + 2wy - v
\]

\section{Applications}

\subsection{The Construction of the Catastrophe Set}

For morphogenetic events that require more structure than elementary catastrophes, the catastrophe set must be constructed. The catastrophe set is composed of elementary (or general) catastrophes that cascade into one another to give rise to complex growth events. Local differential models capture the dynamic in the neighborhood of a catastrophe. The corank of a degenerate critical point determines the minimum spacial parameter necessary to preserve the relevant topology of the dynamic. The dynamic, however, can be embedded in arbitrarily higher spacial dimensions for the purposes of a direct model of embryogenesis. Thom remarks on the theoretical advantage of this property saying,
\begin{center}
{\it The dynamical origin of a catastrophe can be described even when all the internal parameters of the system are not explicitly known. \cite{Thom}}
\end{center}
We consider two examples. The first is a tree without leaves, where the catastrophe set contains only elliptic umbilic catastrophes. The second is a Hamiltonian model with self-reproducing vague attractors and an application of Maxwell's principle.

\subsubsection{Tree without leaves}

Perhaps one of the most elementary applications of this program is the body of a bare tree. Each growth initiation of a branch is an element of the catastrophe set. The structurally stable bifurcation of the dynamic that will generate a branch is the elliptic umbilic $V(u,v,w,x,y)= x^{3} - 3xy^{2} + w(x^{2}+y^{2}) - ux - vy$. The catastrophe set is then composed of elliptic umbilics at locations in the manifold of states that correspond to the topology of tree branches. Examining the gradient model $\dot x = -\nabla V(u,x)$

\[
\dot x =  -3x^{2} + 3y^{2} - 2wx + u
\]

\[
\dot y =  6xy - 2yw + v
\]

If indeed the branch corresponds to the structurally stable elliptic umbilic, the underlying dynamical system of the growth of a tree branch is 2-dimensional, and we expect cell migration to follow a gradient flow in the neighborhood of each element of $K$.
At each new growth event, a thermodynamical coupling can be defined measure theoretically.

The phase space is $(x,y,\dot x,\dot y)$, and the unfolding space (parameter space) is $(u,v,w)$. A catastrophe occurs for certain values of $(u,v,w)$.  
The equilibrium surface is given by

\[
-3x^{2} + 3y^{2} - 2wx + u = 0
\]

\[
6xy - 2yw + v = 0
\]

\subsubsection{Self-Reproducing Singularities}

Thom considers a local Hamiltonian dynamic. Even though Hamiltonian dynamics can't have conventional attractors, they can still have what are known as vague attractors, which we will discuss in more detail in the next section.  Therefore, we can still study a universal unfolding of a singularity as a potential. We will define the Hamiltonian to be $H(p,q) = \frac{p^2}{2m} + V(q).$

\begin{equation}
  \dot q =\frac{\partial H}{\partial p}
  \end{equation}
 
  \begin{equation}
  \dot p =- \frac{\partial H}{\partial q}
\end{equation}

We can examine a codimention three singularity unfolded at the origin. According to Thom,

\[
V_{1} = V(q) + ug_{1}(q) + vg_{2}(q) + wg_{3}(q)
\]

A universal unfolding of this type is what Thom calls a self reproducing singularity.  Self reproducing in the sense that if the singularity $V(x)$ has the potential $V$ within a space of variables $x$ has the same topology type as $V_{2}$.  Configurations with such self reproducing singularities occur in what Thom describes as Maxwell's convention.  In Maxwell's convention one would have a bifurcation set $H$ of the function space $L(U,V)$, which has attractors of a local dynamic corresponding to a minima of the map $g$.  Like before, this bifurcation set $H$ is a strata.  Here the catastrophe points are going to be counter images of the strata with $H^{'}$ of $H$.  Here $H^{'}$ is going to be the set of functions with two local minima.  These strata tend to have a minima that decompose during bifurcation into two distinct minima, hence, self reproducing.  The most straightforward example would be the Riemann-Hugoniot catastrophe $V=x^4/4$, which has the universal unfolding
\begin{equation}
V_{1}=\frac{x^4}{4}-\frac{ux^2}{2}+vx+v^2
\end{equation}
When looking at the equation, there is a curve $x^{4}/4-x^{2}/2$, which gives an even curve containing two symmetrical minima where a curve defined on $V_2$ is going to have the same topology type after applying Maxwell's convention. Maxwell's convention removes all of the critical points except that which has the minimum potential. Multiple critical points may remain if there are several equivalent local minima all lower than their neighbors in a potential well.

\section{Hamiltonian Dynamics}

In this section we continue to carefully follow Thom's original outline in addition to discussing new original mathematical and physical results.  On a differential manifold $M$, the dual of the tangent bundle ($T^{*}M = \cup_{p}T^{*}_{p}M$) has canonical coordinates $(p_{i},q_{i})$ that permits the definition of a symplectic form $\omega = dq^{i} \wedge dp_{i}$, endowing an abstract Hamiltonian flow on $M$. Our field $X_{H}$ is given by the equation $dH = \iota(X_{H}) \cdot \omega$, where $\iota$ is the interior product. In Hamiltonian dynamics there are no attractors specifically due to the invariance of the Liouville measure under the Hamiltonian dynamic.  However, there are specific types of attractors known as vague attractors.  Unlike regular attractors, vague attractors are able to be defined within a Hamiltonian system, and these vague attractors maintain stability under small perturbations of the Hamiltonian.  With ordinary attractors, one can have several basins of attraction competing with one another, and while this is topologically very complicated, one is still able to retain structural stability.  Yet when such a competition between two basins of attractors occurs, the final conditions will be very sensitive to perturbations.  In Hamiltonian dynamics it could be very difficult to speak of basins and you can never be certain regarding the dominance of an attractor.  This is why in the case of vague attractors, the situation is subject to being even more indeterminate, and these vague attractors are going to have stable properties specifically under very small perturbations of the Hamiltonian.  Therefore, such properties are what make the system so indeterminate when working with Hamiltonian systems.  Interestingly enough, these realizations were pointed out by Thom, who stated how this is very reminiscent of the probabilistic nature of quantum mechanics. In return, such properties of vague attractors in the Hamiltonian aren't going to lead to a topological stability, but a global stability of the Hamiltonian.  To begin examining these properties, we can take a field containing the local Hamiltonian dynamics on the open set $W$, which is going to admit the fibration $p:F \rightarrow P$.  From here we can also have the open subset $U$ of $W$ contained in the vague attractor $c(x)$.  If this is the case, we can say the attractor is contained in the fiber of fibration $F \rightarrow P$. Generally there would be a continuum of attractors $c(x)$ within the fiber, where such a continuum would be a sympletic  manifold where a Liouville measure $A$ can be defined.  With this we now have a Liouville measure $A$ and an entropy $S$ pertaining to $c$, which allows us to write out the formula $S=log\: A$.  With the entropy we can now define the temperature $T$ by $T^{-1}=dS/dE$.  Let us now take a group locally containing the same local dynamical equivalence $G$, which is also associated with an attractor $c$.  Here $G$ operates specifically as a subgroup on $W$, and due to $G$ also acting on the continuum of attracts $c(x)$ in the fiber $F$, we are going to have an isotropy group of $G$.  Here we have a transitive relation over the open subset $U$, where in turn we get the dynamic being homogeneous and isotropic in $U$.  Due to these conditions, it would seem we would be able to have a local equilibrium, and if we have an equilibrium, we know that it must not exchange energy with the rest of the system.  Naturally, based on the conditions, we would suppose the system's invariant, where we would have a proportionality between the exchange in energy and the temperature difference resulting in $dE=0$.  Yet it was pointed out by Thom that such conditions are satisfied if and only if $\int_{D} (T - T_{0}) d^{2}x = 0$ on a disc $D$, where $T_{0}$ is the temperature at the center of the disc. Thom gives the condition for equilibrium in terms of the microcanonical entropy. If we define $m(x)$ as the Liouville volume and let $a(x) = \frac{dm}{dx}$, then for two thermally coupled systems $S_{1}$ and $S_{2}$, thermal equilibrium is achieved if  
\begin{equation}
\frac{1}{a_{1}} \frac{d a_{1}}{dt} = \frac{1}{a_{2}} \frac{d a_{2}}{dt}
\end{equation}
where $S = k_{b}$ log$(a(x))$. At equilibrium we see that the temperature acts as a harmonic function $\Delta T=0$, and when not in equilibrium the system will evolve based on $dE/dt=k/\Delta T(E)$.  When acting as a linear oscillator, $T$ and $E$ act as a scalar, which allows us to recover the heat equation.  However, when dealing with quantum fields, $E$ can be complex, which will allow one to recover the Schrodinger equation.  The interesting thought of there being a natural processes that approaches thermal equilibrium through a superposition proposed by Thom, is indeed something well worth investigating further in the future.  Returning to the dynamics of the Hamiltonian systems, we established the existence of a fiber $(M,X)$ over the open set $W$ where there would be a hypersurface pertaining to $E$.  This would allow us to have a finite number of vague attractors $C_1, C_2,..., C_k$, where such vague attractors can contain the corresponding Liouville measures $M_1, M_2,..., M_k$ of the hypersurface $E$.  With the Liouville measures acting as functions on $E$ we can then write the entropy as $S_j=log(m_j)$ and the corresponding temperature $T_j^{-1} = \partial S/\partial E$.  When the phases $X_j$ are of the same temperature, there could be an equilibrium, and there could also be a thermodynamical coupling of the attractors $C_j$.  With these conditions met, this would give rise to a universal thermodynamical coupling of the attractors $C_j$.  It is interesting to note that if the functions $T_j(E)$ are invertible, we would see that $E_j(T)$  would then act as the corresponding energies of the attractors $C_{j}$.  We can take the volumes $V_{1} + V_{2} +,...,V_{K}$ act as the volumes of the corresponding phases described above.  We can then write the phases out as $X_1 + X_2  +,...X_k$, and we then get the following $V_1 + V_2 +...V_K$ = total volume of $W$.  In turn we then can write up $E_{1}V_{1} + E_{2}V_{2} +...E_{K} V_{K}$ = total energy of the system.  While this is a very convenient way to write out and express the total energy of the system, this tells us very little about the volume of the attractors $V_j$ and spatial division of the phase $X_j$.  It would be suspected just like in Maxwell's convention, there is a minimality condition that plays a role. If such a condition plays a role we can have $A_{ij}$ donate the area of a shock wave or perturbation that separates $X_1$ and $X_j$ would have some expression in the form of
\begin{equation}
P = \sum K_{ij} A_{ij} (K_{ij} > 0)
\end{equation}  Such an expression would be expected to be minimized and the topology of the fiber dynamic would seem to have little effect leaving it invariant.  It was suspected by Thom, that many patterns and biological forms such as butterfly wings, shell patterns and formations would be expected to depend on a very similar type of mechanism due to the variety and complexity of the patterns in such formations.

\subsection{Thermodynamic Unfoldings}

While we have shown many different ways to express the Hamiltonian, there is still an elephant in the room, and with the setup we have, that particular elephant begs for attention.  This happens to be the study and formation of a partition function.  We know that a partition function describes a thermodynamical system in equilibrium.  When working out the dynamics of the Hamiltonian we saw that we were able to define a Liouville measure $A$ on a symplectic manifold when examining the vague attractors $c(x)$.  We also saw that there is a continuum of vague attractors that are contained within a fiber where the continuum of the fiber is the corresponding symplectic manifold.  This symplectic manifold was shown that the Liouville measures $A$ are defined.  From the Liouville measure we saw that it was possible to write out the entropy $S=log\: A$ and the temperature as $T_j^{-1} = \partial S/\partial E$.  There also happened to be a subgroup $G$ that had the equivalent dynamics locally, which also operates on the continuum of attractors in addition to the open set $U$ that acts as a subgroup on $W$.  There was an isotropy group of $G$ that was transitive on the open subset $U$ on $W$ where we saw that the corresponding dynamic was isotropic and homogeneous on $U$.  Due to such homogeneity it was easy to write out the exchange of energy in equilibrium expressed as $dE=0$.  As we know, the partition function is used to express the statistical properties in thermodynamical equilibrium.  Taking a Hamiltonian, the partition function can be expressed as
\begin{equation}
Z(\beta)=\sum_{x_{i}}exp(-\beta H(x_1, x_2,...,x_n))
\end{equation} where $H$ is the Hamiltonian, $X_i$ represents a set of random variables and $\beta$ being the inverse temperature.  We have it written as a sum over $X_i$ to express it as the sum over all the possible values the variables $X_i$ could take on.  Yet when examining the local dynamics of the Hamiltonian corresponding to the vague attractors, there is going to be a continuum of attractors $c(x)$.  So for our purposes we can rewrite the partition function as
\begin{equation}
Z(\beta)=\int exp(-\beta H(x_1, x_2,...,x_n))dx_{1}dx_{2}...dx_{n}
\end{equation}
where we simply replace the sum by an integral, making the partition continuous rather than discrete.  Due to Thom's theory being a topologically based theory, it would be most useful to express the partition function canonically in the form
\begin{equation}
Z=\frac{1}{h^3}\int e^{-\beta H(q,p)}d^{3}q d^{3}p
\end{equation}
Canonically, the energies are Boltzmannian distributed \cite{Landau1}. This is a maximization of the Shannon entropy subject to a constant average energy. It is a distribution that is as ``disorganized" as it could be while keeping the average energy constant. A further study of the energy distribution should be done measure theoretically and keeping in mind a thermodynamical coupling between dynamical systems. So we expect a new growth event to, at least provisionally, be thermally coupled to the system that it grew from. These distributions may need to be derived circumstantially.
Proceeding with the canonical partition function, we can stop for a moment and look back at the canonical map which gave the existence of the stratified map
\begin{equation}
F\colon X\times L(X,Y)\rightarrow Y \times L(X,Y)\rightarrow L(X,Y)
\end{equation}
which was defined as
\begin{equation}
(x,f)\rightarrow(y=f(x),f)\rightarrow f
\end{equation}
which was one of the key requirements needed to satisfy the universal unfolding.  We are beginning to see very strong similarities between the universal unfolding and the Hamiltonian dynamics.  Yet this poses the question, is there an analog of the universal unfolding with respect to thermodynamical systems?  The answer seems to be yes.  When analyzing the Hamiltonian dynamics of the fiber $(M,X)$ we saw that the Liouville measures $m$ formed the relative entropy $S_j=log(m_j)$ and the corresponding temperatures $T_j^{-1} = \partial S/\partial E$.  If the phases $X_j$ of the system are the same temperature there will be thermodynamical equilibrium $dE = 0$, and if we have thermodynamical equilibrium we will in turn have a universal coupling between the attractors $c_j$.  Such attractors have the Liouville measure that form the entropy $S_j=log(m_j)$ and temperature $T_j^{-1} = \partial S/\partial E$.  It was shown that the phases $X_j$ controlled by the attractors are required to be in thermal equilibrium.  In return this shows the manifestation of a canonical partition function from a purely geometric theory.  There is one other key component to which we should turn our attention: the unfolding of the singularities of the Hamiltonian. Luckily we already have a fibration $F\rightarrow P$.  Having a fibration makes it possible to study the mapping of the singular points of the symplectic manifold that is formed by the continuum of attractors $c(x)$.  Since we have a continuum and have the existence of a fiber bundle, this tells us that the mapping is going to be a surjective submersion.  Having surjective properties would tell us that we can have a function $f$ where we can take the domain $X$ and codomain $Y$, where $X$ would have elements $x$ and $Y$ containing elements $y$.  Also due to there being surjectivity, $Y$ would contain elements $y$, and for every element of $y$ in $Y$, there would be at least one element $x$ in $X$.  So we would get the function $f(x)=y$.  In turn, having a submersion tells us that we would have a map between two differentiable manifolds $N$ and $M$ where the differentiable is everywhere surjective.  Specifically, we can take the map $f:M\rightarrow N$ where the map $f$ would have a submersion at a point $p$ which would be an element of $M$.  Therefore taking the differential of the function at the point would give \begin{equation}
Df_p:T_pM\rightarrow T_{f(p)}N
\end{equation}
For the canonical map
\begin{equation}
F\colon X\times L(X,Y)\rightarrow Y \times L(X,Y)\rightarrow L(X,Y)
\end{equation}
we had a very similar setup with two differential manifolds $X$ and $Y$ which were expressed as the function space $L(X,Y)$.  Yet looking back at our solution, we saw that the canonical map was defined as \begin{equation}
(x,f)\rightarrow(y=f(x),f)\rightarrow f
\end{equation}
If we carefully look at the above equation, we are able to see that there is a surjective function $f(x)=y$.  Yet we have also shown that the canonical map gives the existence of a stratified map, which was needed to satisfy the universal unfolding.  When we turned our attention to the vague attractors of the Hamiltonian dynamics there was the emergence of a fibration.  The fibration allowed us to examine the mapping of the points on the symplectic manifold.  However, what we would really be interested in are more specifically the singularities.  Therefore, we would be interested in a map that is a fibration that would allow us to study said singularities.  Such a map would be a Milnor map, which is also known as a Milnor fibration.  To define a Milnor map we can take $f(Z_0,...,Z_n)$ representing a polynomial function of complex variables, where the vanishing of a locus would be at the origin.  One can also have a Milnor fiber of an isolated hypersurface containing singularities.  So taking the singularity at the origin, we would have $f=0$.  It is important to state that the polynomial function $f(Z_0,...,Z_n)$ is going to contain $n+1$ complex variables.  So we are going to have a $\mathbb{C}^{n+1}$ topology.  So we will have $f_t:\mathbb{C}^{n+1}\rightarrow \mathbb{C}$ where the polynomial becomes $(Z_0,...,Z_n)\rightarrow f(Z_0,...,Z_n)-t$, where $f=0$ is now $f_t\neq 0$.  Here we have a polynomial that had a singularity at the origin, which went under a type of unfolding and became a polynomial.  We saw extremely similar situations when analyzing the jet space.  Recall that we began studying the jet space $z$ by analyzing the problem locally where we had the germ $g\colon \mathbb{R}^m \rightarrow \mathbb{R}^n$ of a differential map, which had the corresponding origin $O$.  The germ allowed one to send the origin $O$ in $\mathbb{R}^m$ to the origin $O_1$ in $\mathbb{R}^n$, which gave the expansion of the map.  From there we looked at the jet space that defined the expansion.  Let us recall that such an expansion allowed us to define the algebraic subset, which was our bifurcation set $K$.  Seeing how the topological space containing the germ was invariant under deformations, rather than just focus on the Milnor maps, we can now look at Milnor numbers.  The advantage of Milnor numbers is that in singularity theory, they are the invariant of a germ $f$ and it is a complex-valued holomorphic function that is non-negative or infinite.  This can be defined as either a geometric or algebraic invariant where there is once again a hypersurface singularity.  A germ of the kind can simply be defined as $f:(\mathbb{C}^{n},0)\rightarrow (\mathbb{C},0)$. When we looked at the field of the Hamiltonian dynamics, we saw that there was a local field of the Hamiltonian.  In return, we also had a fiber that indeed had a hypersurface, which had the finite number of vague attractors, that formed the relative entropies and attractors.  We then saw that if the phases were the same temperature there would be equilibrium, which would give us a universal thermodynamical coupling.  The conditions for the universal thermodynamic coupling satisfied the conditions for us to derive the partition function.  However, we now see that this allows us to derive much more.  With the universal unfolding, we had the singularities $\mathbb{R}^m\rightarrow \mathbb{R}^q$ pertain to the universal map $\mathbb{R}^m\rightarrow \mathbb{R}^q$.  The formation of the universal map began with the germ that gave the expansion of the singularities. This then gave the formation of the jet space that defined the expansion.  In the Hamiltonian dynamics there exists the universal thermodynamical coupling with the fibration that allowed us to study the mapping of singular points on the symplectic manifold.  On the sympletic manifold there existed the continuum of vague attractors, which gave us our fiber bundle.  From the fiber bundle we had the surjective submersion where we took the mapping of the two differentiable manifolds $M\rightarrow N$, which simply gave rise to the solution of the canonical map that had the surjective function $f(x)=y$.  This also satisfies the conditions for a stratified map, which had the subspace of the closed bifurcation set.  The very proof of such a stratified space was the study of the universal family, which arose from the germ of the differential map that ending up giving topological invariance.  Thanks to the existence of the fibration of the Hamiltonian, we were able to replicate this by taking both the Milnor map and Milnor numbers.

The Milnor map or the Milnor fibration is a specific type of fibration that is used to study singularities.  We saw that the mapping of such singularities of a differential function that gave the emergence of a polynomial in the same way when studying the properties of the universal unfolding.  The Milnor map that satisfies the properties of the universal unfolding also satisfied the properties of the universal thermodynamic coupling due to being a fibration of the hypersurface containing the vague attractors.  From the analysis of the Milnor maps we saw that the Milnor numbers; just like the formation of the universal unfolding, they are the invariant of a germ which shows us that there is a geometric invariance.  Hence, we have shown that the universal thermodynamical coupling satisfies the conditions of the universal unfolding.  Therefore, the universal thermodynamical coupling gives rise to the Hamiltonian that is an analog of the universal unfolding.

\section{Conclusion}

Thom's morphogenetic models remain open to investigation in an epoch of optical experimental techniques that make possible precise measurements of local embryological development. A reexamination of these methods by the scientific community may be timely.

While Thom's approach doesn't offer a direct quantitative model, it does outline a unified approach to reconstructing the phase space profile. This means that experimental verification must be done in tandem with modern techniques in data science. Empirically, and for appropriately small distances and times, the trajectory of cell migration must be tracked until a phase space profile can be reconstructed. We can verify that the underlying dynamic of morphogenesis is indeed determined by the structural stability of topological dynamical systems permitted by the structure of our spacetime, if the empirical phase space profile is diffeomorphic to Thom's respective local models. Thom's local models, however, are phenomenological; that is, upon observing the initiation of a new growth event, a catastrophe is selected appropriate to the final form that the event will produce. In that sense, the theory is finalistic (i.e. we must already know the final state to apply it). However, the selection of a final state is not as free as Lagrangian mechanics. The final states are already classified and determined by the dimension and topology of space. With this framework, we can begin to address larger concerns. This classification of morphologies within Thom's framework defines the question: {\it what can be grown in our universe?} 

Furthermore, if structural stability is indeed the final principle in these developmental events, the existence of low dimensional models in otherwise high dimensional data is verified. This represents an encouraging step forward in our search for simple models of biophysical processes. To what extend a complete theory of morphogenesis can be deduced from more fundamental principles remains on open question, and theoretical approaches to biology remain almost as diverse as the phenomena they wish to describe \cite{Reed}. Unlike theoretical physics, where complexity, in a gas or a solid for example, is treated statistically and therefore reductionist, the open door to biological theory may be through finding simplicity in the final product of biological interactions. We then use that framework to construct theories of constituent complex behavior. 

We have shown and reviewed the mathematical framework of Thom's theory of morphogenesis.  We went over necessary requirements needed to be made for the theory to emerge.  From the mathematical framework we also went over the emergence of the physical applications that manifest.  Such applications appear to give rise to new realizations and insights into areas such as Hamiltonian mechanics and thermodynamics.  These physical applications have also been shown to be dual to a purely geometric and topological theory.  Much investigation is still required, and many new avenues have been opened for one to explore. We expect these tools to be useful not only in the direct modeling of biological development, but also in the motivation of new theoretical methods.  At the very least we can conclude that there are exciting new mathematical and physical realizations to be discovered in the future.

\section*{Acknowledgements}

We thank Prof. Robert Alfano (Institute for Ultrafast Spectroscopy and Lasers of the City University of New York) for discussing the content and the presentation of this work. We thank Johannes Zwanikken (University of Massachusetts Lowell), Rahul Kashyap (Penn State University) and Christopher Bresten (Ajou University, Suwon, Korea) for helpful discussions. We thank Chris Peter (University of Massachusetts Dartmouth) for help in the professional editing of this material.  We also thank Mailde S. Ozório (University of São Paulo) for help in the final edit of our manuscript.

\section*{Appendix}

To further realize Thom's original program, a deeper look at the partition function is needed. A discussion of the search for lower dimensional behavior within biological systems is done in \cite{Bialek}, and the author contextualizes the role of dimensionality in modeling biological systems. A probability distribution over trajectories is given by

\[
P[x(t)] = \frac{1}{Z}e^{-S[x(t)]}
\]

for a dynamical system with a Gaussian noise term. Consider the gradient dynamic for invagination

\[
\dot x = -x^4 - ux^2 - vx - w
\]

Since this dynamical system is by derivation structurally stable, it is diffeomorphic to

\[
\dot x = -x^4 - ux^2 - vx - w + \eta(t)
\]

where $\eta(t)$ is a Gaussian noise term. $S[x(t)]$ is defined as

\[
S[x(t)] = \frac{1}{2} \int \int x(t)K(t- t')x(t') dt dt'
\]

The Gaussian kernel is

\[
K(t + t') = \sum^{D}_{n=1} a_{n}\phi_{n}(t)\phi_{n}(t')
\]

Hidden dimensions are identified according to the structure of correlations in the noise source. The dimension $D$ of the dynamical system specifies the number of initial conditions needed to determine the system.  While a partition function of this kind may not have the topological structure necessary to satisfy the universal unfolding, the use of a partition function of this kind may be useful when examining the physical applications.  These applications would include the thermodynamic properties of the system, such as probability distributions and the entropy of the system.


\begin{thebibliography}{0}%
\makeatletter
\providecommand \@ifxundefined [1]{%
 \@ifx{#1\undefined}
}%
\providecommand \@ifnum [1]{%
 \ifnum #1\expandafter \@firstoftwo
 \else \expandafter \@secondoftwo
 \fi
}%
\providecommand \@ifx [1]{%
 \ifx #1\expandafter \@firstoftwo
 \else \expandafter \@secondoftwo
 \fi
}%
\providecommand \natexlab [1]{#1}%
\providecommand \enquote  [1]{``#1''}%
\providecommand \bibnamefont  [1]{#1}%
\providecommand \bibfnamefont [1]{#1}%
\providecommand \citenamefont [1]{#1}%
\providecommand \href@noop [0]{\@secondoftwo}%
\providecommand \href [0]{\begingroup \@sanitize@url \@href}%
\providecommand \@href[1]{\@@startlink{#1}\@@href}%
\providecommand \@@href[1]{\endgroup#1\@@endlink}%
\providecommand \@sanitize@url [0]{\catcode `\\12\catcode `\$12\catcode
  `\&12\catcode `\#12\catcode `\^12\catcode `\_12\catcode `\%12\relax}%
\providecommand \@@startlink[1]{}%
\providecommand \@@endlink[0]{}%
\providecommand \url  [0]{\begingroup\@sanitize@url \@url }%
\providecommand \@url [1]{\endgroup\@href {#1}{\urlprefix }}%
\providecommand \urlprefix  [0]{URL }%
\providecommand \Eprint [0]{\href }%
\providecommand \doibase [0]{http://dx.doi.org/}%
\providecommand \selectlanguage [0]{\@gobble}%
\providecommand \bibinfo  [0]{\@secondoftwo}%
\providecommand \bibfield  [0]{\@secondoftwo}%
\providecommand \translation [1]{[#1]}%
\providecommand \BibitemOpen [0]{}%
\providecommand \bibitemStop [0]{}%
\providecommand \bibitemNoStop [0]{.\EOS\space}%
\providecommand \EOS [0]{\spacefactor3000\relax}%
\providecommand \BibitemShut  [1]{\csname bibitem#1\endcsname}%
\let\auto@bib@innerbib\@empty
\end{thebibliography}%


\newpage

\begin{thebibliography}{9}
\bibitem{Thom}
Thom, Ren\'{e}.
{\it Structural Stability and Morphogenesis: A General Theory of Models.}
Thom, R. (1975). Structural stability and Morphogenesis: An outline of a general theory of models. Reading: Benjamin.
\bibitem{Bialek}
W Bialek, What do we mean by the dimensionality of behavior? arXiv:2008.09574 [q–bio.NC] (2020).
\bibitem{Landau1}
L. D. Landau, E. M. Lifshitz, Statistical Physics, Third Edition, Part 1, Pergamon Press, Oxford (1980).
\bibitem{Reed}
M. Reed, Why Is Mathematical Biology So Hard? Notices of the American Mathematical Society (2004).





\end{thebibliography}
\end{document}